\documentclass[twocolumn,showpacs,amsmath,amssymb,prd,aps]{revtex4}
\usepackage{graphicx}
\usepackage{epstopdf}

\usepackage{amssymb}
\usepackage{amsmath}
\usepackage{graphicx}
\usepackage{epstopdf}
\usepackage{color}

\newcommand{\req}[1]{Eq.\,(\ref{#1})}

\newcommand{\MeV}{\text{ MeV}}

\newcommand{\Neff}{N_{\text{eff}}}

\begin{document}

\title{Quark-Gluon Plasma as the Possible Source of Cosmological Dark Radiation}
\author{Jeremiah Birrell and Johann Rafelski}
\address{Department of Physics, The University of Arizona,  Tucson  Arizona, 85721, USA}

\date{November 22, 2014}

\begin{abstract}
The effective number  of neutrinos, $\Neff$, obtained from CMB fluctuations accounts for all effectively  massless degrees of freedom present in the Universe,  including but not limited to the three known neutrinos.  Using  a lattice-QCD derived  QGP  equation of state, we constrain  the observed range of $\Neff$  in terms of the  the freeze-out of unknown degrees of freedom near to quark-gluon hadronization. We explore  limits on the coupling of these particles, applying methods of kinetic theory. We present bounds on the coupling of such  particles and discuss the implications of a connection between $\Neff$ and the QGP transformation for laboratory studies of QGP.
\end{abstract}


\maketitle
\noindent{\bf Introduction:} Phase transitions in the early Universe,  such as the Electroweak and QCD transitions, constitute a drastic change in the properties of the vacuum.  In the case of QCD, the strong symmetry  breaking is accompanied by the presence of relatively heavy Goldstone bosons. It is natural to wonder whether such a transition comprises further and much weaker symmetry breaking, accompanied by low mass (sub eV scale mass) Goldstone bosons, expected to decouple at or near  the phase boundary. We refer to any light weakly coupled  particle species as a `sterile particle' (SP),   generalizing the  sterile neutrino concept-- to avoid misunderstandings we stress that these SPs are not (cold) dark matter  but rather behave as `dark radiation'~\cite{Steigman:2013yua}.

To motivate the assumption that the transformation of the vacuum structure is the origin of SPs, it is best to compare our discussion with Ref.~\cite{Weinberg:2013kea}, where a concrete but yet to-be-discovered model is proposed.  In contrast, here we consider the latest  phase transformation in the early Universe and evaluate quantitatively  the production and freeze-out of possible SPs in such a transition. If SPs are interpreted as Goldstone bosons, it would imply that in the deconfined phase there is  an additional hidden symmetry, weakly broken at hadronization.  For example, if this symmetry were to be part of the baryon conservation riddle, then we can expect that these Goldstone bosons will  couple to particles with baryon number, and possibly only in the domain where the vacuum is modified from its present day condition. 

Another viable candidate for SPs are sterile neutrinos. It was shown that the freeze-out temperature required for three `new' right-handed neutrinos to fully account for the effective number of neutrinos, $\Neff$  (see the following section),  is in the vicinity of the quark gluon plasma (QGP) phase transition~\cite{Anchordoqui:2011nh,Anchordoqui:2012qu}.  The former paper proposed a concrete model of how this might be obtained from an expanded gauge group for QCD.  However,  the QGP equation of state (EoS) which were used are not consistent with recent numerical lattice-QCD results.   We use the  lattice-QCD derived QGP EoS from Ref.~\cite{Borsanyi:2013bia} to characterize the relation between $\Neff$ and the number of DoF that froze out at the time that the quark-gluon deconfined phase froze into hadrons near $T=150\MeV$, and  compute the coupling strength required to achieve freeze-out at the QGP transformation. 

The question of the validity of the symmetry breaking model~\cite{Anchordoqui:2011nh}  is far from resolved, and other extensions of strong interactions   are present in the literature, see for example \cite{Georgi:1989gp}. The cancellation seen in lattice simulations between QED and QCD CP-odd terms~\cite{Bonati:2013hsa,D'Elia:2012zw}  provides concrete evidence of a possible connection between the QED and QCD sectors, showing that the theory of strong interactions is not fully understood and can contain symmetry breaking outside the known realm.

Even a very weakly coupled SP, should it be associated with symmetry breaking below and at QGP hadronization, could be seen in several experiments, a point we develop in this work:\\
a)  An experimental motivation for prior interest is the analysis of CMB temperature fluctuations, such as by the Planck satellite collaboration (Planck)~\cite{Planck}, especially the observed tension in the effective number of neutrinos, $\Neff$. $\Neff$ is constrained by the expansion of the Universe and includes all light sub-eV mass particles present in the Universe, such as our SPs.\\
b)   By applying methods of kinetic theory, we obtain the minimal coupling strength of SPs required to maintain chemical equilibrium down to the time of  hadronization of cosmological or laboratory formed QGP.  We show that in this situation a significant fraction of the total energy of the QGP phase could be unaccounted for, carried out by SPs  that are `invisible'  in the confined vacuum.  This result motivates a closer study of the energy balance and expansion dynamics in laboratory QGP experiments.

Our discussion of the role of SPs in understanding the early Universe expansion complements and competes with other explanations of the tension in  $\Neff$, which has already inspired various theories, including the consideration of:
i)  a model in which the temperature of neutrino decoupling was a variable parameter \cite{Birrell:2013_2};
ii) a very light neutralino that freezes-out prior to muon annihilation~\cite{Dreiner:2011fp};
iii) as noted, the introduction of Goldstone bosons associated with a new spontaneously broken symmetry that freeze out prior to the disappearance of muons, making a fractional contribution to $\Neff$~\cite{Weinberg:2013kea}. 
 
The last case  is an example of the general mechanism whereby ultra-weakly interacting particles of any type that freeze-out in an earlier epoch of the Universe, such as our SPs, make a contribution to $\Neff$ that depends on the decoupling temperature~\cite{Anchordoqui:2011nh,Anchordoqui:2012qu,Blennow:2012de,Steigman:2013yua}. This  naturally results in a fractional contribution to $\Neff$. After decoupling, SPs do not participate in the reheating process, in which the entropy of a disappearing  particle component is transferred into the remaining components. The noticeably lower temperature of SPs, compared to the reference particle (photon), means they have a smaller contribution to thermal pressure and energy, an effect measured by $\Neff$, resulting in a fractional contribution to the `neutrino' DoF. \\[-0.2cm]

\noindent{\bf Effective Number of Neutrinos:}
$\Neff$ quantifies the amount of radiation energy density, $\rho_r$, in the Universe prior to photon freeze-out and after $e^\pm$ annihilation and is defined by $\rho_r=(1+(7/8)R_\nu^{4}\Neff)\rho_\gamma$, where $\rho_\gamma$ is the photon energy density and  $R_\nu\equiv T_\nu/T_\gamma=({4}/{11})^{1/3}$ is the photon to neutrino temperature ratio in the limit where no entropy from the annihilating $e^\pm$ pairs is transferred to neutrinos.  The factor 7/8 is the ratio of Fermi to Bose reference normalization in $\rho$ and the neutrino to photon temperature ratio $R_\nu$ is the result of the transfer of $e^\pm$ entropy into photons after Standard Model (SM) left handed neutrino freeze-out.

If photons and SM left-handed neutrinos are the only significant massless particle species in the Universe between the freeze-out of left-handed neutrinos at  $T_\gamma=\mathcal{O}(1)$ MeV and photon freeze-out at $T_\gamma=0.25$ eV, and assuming zero reheating of neutrinos, then $\Neff=3$, corresponding to the number of SM neutrino flavors  by definition.  A numerical computation of the neutrino freeze-out process employing SM two body scattering interactions and carried out using the Boltzmann equation framework presented in~\cite{Dicus:1982bz} gives $\Neff^{\rm th}=3.046$~\cite{Mangano2005}, a value close to the number of flavors.  

The value of $\Neff$ can be  measured by fitting to observational data, such as the distribution of CMB temperature fluctuations. The  Planck~\cite{Planck}  analysis gives $\Neff=3.36\pm 0.34$ (CMB only) and $\Neff=3.62\pm 0.25$ (CMB+$H_0$) ($68\%$ confidence levels). With more dedicated CMB experiments forthcoming and an analysis that can self consistently account for any additional particle inventory in the early universe, it is believed that a significantly more precise value of $\Neff$  will be available in the next decade. \\[-0.2cm]

\noindent{\bf Contribution to $\Neff$ of a Sub-eV  Mass SPs:}  The Einstein equations  imply a practically entropy conserving expansion of the Universe. Entropy conservation during periods when dimensional (mass) scales are irrelevant means that all temperatures scale inversely with the metric  scale factor $a(t)$. As temperature  passes through $m\simeq T$ thresholds, successively less massive particles annihilate and their entropy is shifted into the remaining effectively massless particles, causing the  $T\propto 1/a(t)$ scaling to break down.

After an effectively massless particle species decouples, its temperature scales as $1/a(t)$ at all later times as a result of the free-streaming solution of the Einstein-Vlasov equation.  This leads to a temperature difference between the free streaming particles, and the photon background, which is the last to freeze-out. This reheating effect builds up during each  period  in which particle species disappear from the Universe inventory.

We denote by $S$ the conserved `comoving' entropy in a volume element $dV$, which scales with the factor $a(t)^3$. We define the effective number of entropy DoF, $g_*^S$, by
\begin{equation}
S=\frac{2\pi^2}{45}g^S_*T_\gamma^3 a^3.
\end{equation} 
For ideal Fermi and Bose gases
\begin{equation}
g_*^S=\!\!\!\!\sum_{i=\text{bosons}}\!\!\!\!g_i \left(\frac{T_i}{T_\gamma}\right)^3\!\!\!f_i^-+\frac{7}{8}\!\!\!\sum_{i=\text{fermions}}\!\!\!\! g_i \left(\frac{T_i}{T_\gamma}\right)^3\!\!\!f_i^+.
\end{equation}
The $g_i$ are degeneracies, $f_i^\pm\in (0,1)$ are known  functions that turn off the various species as the temperature drops below their mass-- compare to the analogous Eqs. (2.3) and (2.4) in Ref.\cite{Blennow:2012de}.

\begin{figure} 
\centering
\begin{minipage}[b]{.49\textwidth}
\centerline{\hspace*{-0.10cm}\includegraphics[height=6.4cm]{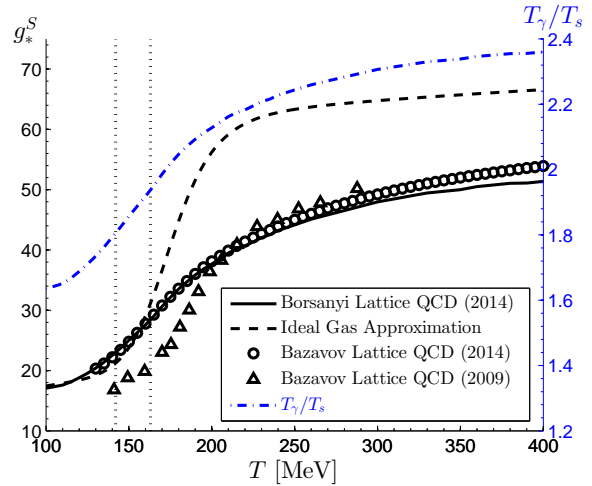}}
\end{minipage}
\caption{Left axis: Effective number of entropy-DoF, including lattice QCD effects applying Ref.~\cite{Borsanyi:2013bia} (solid line) and  Ref.~\cite{Bazavov:2014pvz} (circles), compared to the early Ref.\cite{Bazavov:2009zn} (triangles) results used by~\cite{Anchordoqui:2011nh}, and the ideal gas model of Ref.\cite{Coleman:2003hs} (dashed line) as function of temperature $T$. Right axis: Photon to SP temperature ratio, $T_\gamma/T_s$, as a function of SP decoupling temperature (dash-dotted (blue) line). The vertical dotted lines at $T=142$ and 163 MeV delimit the QGP transformation region.\label{fig:gS}}
 \end{figure}

Such a simple characterization does not hold in the vicinity of the QGP phase transformation where  quark-hadron degrees of freedom are strongly coupled  and the system must be studied using lattice QCD. This result is incorporated in the solid line in figure \ref{fig:gS} (left axis), where we have used a table of entropy density values through the QGP phase transition presented by Borsanyi et al.~\cite{Borsanyi:2013bia}, while circles show recent results from Bazavov et al.~\cite{Bazavov:2014pvz}. This should be compared to the ideal gas approximation from~\cite{Coleman:2003hs} together with the fit in~\cite{Wantz:2009it} to interpolate though the QGP phase transition and older (year 2009) lattice data from Ref.\cite{Bazavov:2009zn} (triangles). The free gas approximation carries with it a maximum error of $10\%$ in the temperature range of the QGP phase transition  $T\simeq 150$\,MeV where quarks appear.  The error in the 2009 lattice data used in Ref.\cite{Anchordoqui:2011nh} is on the order of $25\%$.  This leads to a non-negligible difference in the relation between freeze-out temperature and $\Neff$.

\begin{figure}
\centering
\begin{minipage}[b]{.49\textwidth}
\centerline{\hspace*{0.4cm}\includegraphics[height=6.8cm]{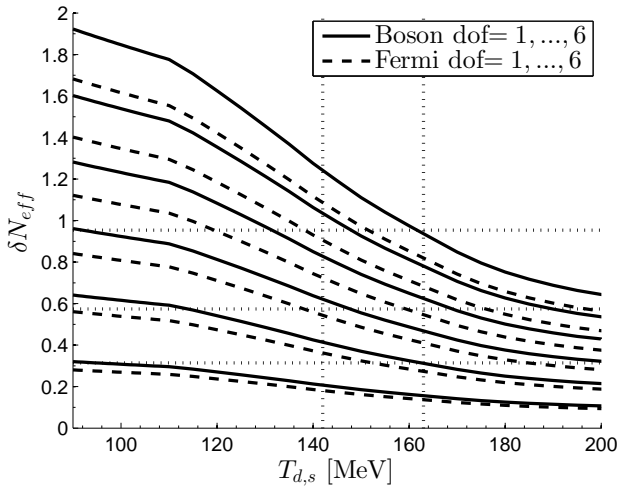}}
\end{minipage}
\caption{Solid lines: Increase in $\delta\Neff$ due to the effect of $1,\dots,6$ light sterile boson DoF ($g_s=1,\dots,6$, bottom to top curves) as a function of freeze-out temperature $T_{d,s}$. Dashed lines: Increase in $\delta\Neff$ due to the effect of $1,\dots,6$ light sterile fermion DoF ($g_s=7/8\times 1,\dots,7/8\times 6$, bottom to top curves) as a function of freeze-out temperature $T_{d,s}$. The horizontal dotted lines correspond to $\delta\Neff+0.046=0.36,0.62,1$. The vertical dotted lines show the reported range of QGP transformation temperatures $T_c=142-163\MeV$.\label{fig:Neff_Td_zoom}}
\end{figure}

Once the SPs decouple from the  particle inventory at a photon temperature of $T_{d,s}$, a difference in their temperature from that of photons will build up during subsequent photon reheating periods, as discussed above. Conservation of entropy leads to a temperature ratio at $T_\gamma<T_{d,s}$, shown in the dot-dashed line in figure \ref{fig:gS} (right axis), of
\begin{equation}\label{T_ratio}
R_s\equiv T_{s}/T_{\gamma}=\left(\frac{g_*^S(T_\gamma)}{g_*^S(T_{d,s})}\right)^{1/3}.
\end{equation}

If $T_s$ and $T_\gamma$ are the light SP and photon temperatures, both after $e^\pm$ annihilation, and $g_s$ is the number of DoF of the SPs normalized to bosons (i.e. for fermions it includes an additional factor of $7/8$) then this gives
\begin{equation}\label{Neff1}
\delta \Neff\equiv \Neff-3.046=\frac{4g_s}{7}\left(\frac{T_s}{R_s T_{\gamma}}\right)^4
\end{equation}
where $3.046$ is the SM neutrino contribution. Using \req{T_ratio} we can rewrite $\delta \Neff$ as
\begin{equation}\label{delta_N}
\delta \Neff=\frac{4g_s}{7R_\nu^4}\left(\frac{g_*^S(T_{\gamma})}{g_*^S(T_{d,s})}\right)^{4/3}.
\end{equation}
where   $T_{d,s}$ is the decoupling temperature of the SP and $T_{\gamma}$ is any photon temperature $T_{\gamma}\ll m_e$. The SM particles remaining at $T_{\gamma}$ are  photons and SM neutrinos, the latter with temperature $R_\nu T_{\gamma}$, and so $g_*^S(T_{\gamma})=2+7/8\times 6\times 4/11$ and (see also Eq.(2.7) in~\cite{Blennow:2012de})
\begin{align}\label{delta_N2}
\delta \Neff\approx&g_s\left(\frac{7.06}{g_*^S(T_{d,s})}\right)^{4/3}.
\end{align}

Figure \ref{fig:Neff_Td_zoom} shows $\delta \Neff$ as a function of $T_{d,s}$ for $1,\dots,6$ boson (solid lines) and fermion (dashed lines) DoF. For a low decoupling temperature $T_{d,s}<100$\,MeV  a single bose or fermi SP  can help alleviate the tension in $\Neff$. Within QGP hadronization interval $T_c=142-163\MeV$ (marked by vertical lines), where SPs could decouple, we see that three bose degrees of freedom or four fermi degrees of freedom are the most likely cases to resolve the tension. 

It is also clear from  figure \ref{fig:Neff_Td_zoom} that the rapid growth of the number of degrees of freedom in the QGP phase implies that earlier decoupling temperatures lead to a rapid increase in the required number of SPs. While one cannot exclude the possible presence of 20--30 new dark light particles, it seems to us unlikely that there are that many undiscovered weakly broken symmetries producing light Goldstone, or/and sterile neutrino-like particles. We believe that figure \ref{fig:Neff_Td_zoom}  pinpoints the QGP  temperature range and below as the primary domain of interest for the freeze-out of such hypothetical degrees of freedom, should these be responsible for the modification $\delta \Neff$.\\[-0.2cm]

{\bf Chemical Equilibration of SPs:}
Using the method for computing freeze-out temperatures via Boltzmann two body scattering operators presented in \cite{Birrell:2014uka}, we can estimate the minimum coupling required within the QGP phase to maintain chemical equilibrium, both in a cosmological and a laboratory setting.  At the present level of precision, we find it reasonable to take an illustrative model, wherein we think of the SPs as if they were sterile neutrinos and model their  interaction and  freeze-out in a manner similar that of the left-handed neutrinos at $T_f=\mathcal{O}(1\MeV)$~\cite{Mangano2005}.  However, in the QGP phase with $T_f\simeq 150$\,MeV, in addition to leptons we also include  coupling to the more abundant quarks. To account for quarks, in this first estimate of the coupling strength we have effectively increased  the number of active degrees of freedom in our computation by the appropriate amount.  In such a model of SP freeze-out, the strength of the interaction is controlled by a modified Fermi constant $  G_{SP}=C G_F$. 

In a cosmological setting, the lower bound for $C$ that will assure that SPs remain in chemical equilibrium until the confining QGP transition into regular matter at $T=\mathcal{O}(150\MeV)$ is approximately
\begin{equation}
C\gtrsim 2\, 10^{-3}, \qquad   G_{SP}^{-1/2}\lesssim 6.5\,\mathrm{TeV}.
\end{equation}
This large $\mathrm{TeV}$  energy scale  for the coupling of e.g. sterile neutrinos seems reasonable and renders such particles within a range that can perturb experimental laboratory data. This result justifies the implementation of a more concrete sterile neutrino coupling model in the early Universe to set a better limit on the model  parameters.

A much greater coupling $C$ is required to assure that a chemical equilibrium abundance of SPs is achieved in the short lifespan of QGP formed in laboratory  heavy ion collisions. In order to model this situation we need a   temperature profile. We assume a simple model $T\tau =T_0\tau_0$. We choose  $\tau_0=8\, 10^{-24}$s  so that the temperature  falls from $T_0=600\MeV$ to $T=150\MeV$ in   $2.4\cdot 10^{-23}$s.  With this, the limit on the required coupling is
\begin{equation}
C\gtrsim 2\, 10^6. \quad \quad   G_{SP}^{-1/2}\lesssim  200\,\mathrm{MeV}.
\end{equation}
The appearance of a coupling on the order of the QCD scale  is consistent with the intuition about the interaction strength that is required for   particles to reach chemical equilibrium in laboratory  QGP experiments.\\[-0.2cm]

{\bf QGP Signature of SP Production}:  As we have seen, for QCD-scale coupling, SPs associated with the deconfined phase transition could be produced abundantly in laboratory relativistic heavy ion experiments, saturating the volume occupied by the QGP with their practically massless yield.  However, to be consistent with the present day invisibility of SPs, their interaction with other particles must only turn on in the domain where the vacuum is modified at finite temperature, analogous to the enhancement of anomalous baryon-number non-conservation at GUT scale temperatures~\cite{Kuzmin:1985mm}.  Therefore, in our proposed scenario, the QGP transition must be associated with a sharp cutoff of the coupling, and thus  scattering cross section,  of the SPs.

Reinspecting the results we have presented,  in particular the  figures, we note the best $\Neff$ constraint suggests a multiplicity of $3\pm 2$ SPs at QGP hadronization,  at which point there are about $g_*^S=25$ strong interaction (entropy) degrees of freedom. This means that approximately $12\pm 8\%$ of all entropy content of the QGP  is within the escaping SPs.  Moreover, since SP's stop interacting at the QGP surface they can escape freely during the entire lifespan of the QGP. As a consequence, the energy loss could be even greater.

We recognize  that the likely loss of energy and entropy could be substantial. A full model of this dynamical process is beyond the present discussion.  However, a quarter or more of the energy brought in by heavy ions into the space-time domain could literally `evaporate'. A systematic exploration of the thermal energy in the QGP fireball  is presented in tables 8 and 9 in Ref.\cite{Letessier:2005qe}. This study did not consider the kinetic energy due to collective matter flow.  However, near   the QGP formation threshold   the  kinetic flow energy component should be small. Inspecting the fireball thermal energy per baryon content near this threshold, we note missing energy of the  here estimated magnitude:   only 75\% of the energy per baryon is found in the visible QGP reaction products.

There is another, indirect, signature of the SPs should they be strongly coupled only within the domain of the QGP phase. We recall that lattice-QCD results predict a relatively small continuous pressure for the QGP phase near the transformation, a situation consistent with absence of a phase transition~\cite{Borsanyi:2013bia}. However, SPs contribute to the pressure internal to QGP, scattering from QGP partons, yet  not in the external region. Therefore, by including SPs there is now a pressure discontinuity at the QGP surface.  This restores the appearance of a phase transition between the deconfined and confined domains and drives the QGP expansion.  In this way, SPs could also be observed through their indirect, dynamical effect on the flow of matter, including particle $v_2$, the  dynamical azimuthal asphericity~\cite{Ollitrault:1992bk,Voloshin:1994mz},  imparting this effect on all components of the QGP, including heavy quarks.\\[-0.2cm]

\noindent{\bf Discussion:} 
The natural concordance of the reported CMB range of $\Neff$ with the range of QGP hadronization temperatures, as seen in figure \ref{fig:Neff_Td_zoom}, motivates the exploration of a connection between $\Neff$ and the decoupling of SPs at and below  the QGP phase transition. We estimated the minimal strength of the coupling of  novel particles such as sterile neutrinos in the early Universe. We further considered the possibility of SPs that are coupled much more strongly, but exclusively within the context of the deconfined vacuum structure.  Under this hypothesis the strength of the coupling could be governed by the QCD scale. We argued that such SPs, possibly a novel type of Goldstone bosons, would have considerable observable impact on both energy balance in the formation of the QGP phase, and on its dynamical evolution.

In summary, we can say that   $\Neff>3.05$ can be associated  with the appearance of several  (best fit a total of $3\pm2$) light  particles at QGP hadronization in the early Universe that either are weakly interacting in the entire space or are allowed to interact only inside the deconfined domain, in which case their coupling would be strong. Such particles could leave a clear experimental signature in, for example, relativistic heavy ion experiments that produce the deconfined QGP phase.

\noindent{\bf Acknowledgments} 
This work was conducted with Government support under and awarded by DoD, Air Force Office of Scientific Research, National Defense Science and Engineering Graduate (NDSEG) Fellowship, 32 CFR 168a, and it has been supported by a grant from the U.S. Department of Energy, DE-FG02-04ER41318.

\bibliography{refs} 

\end{document}